\documentclass[preprint,journal]{vgtc}       

\usepackage{booktabs} 
\usepackage[colorinlistoftodos]{todonotes}

\usepackage{makecell}   

\usepackage{amsmath}
\usepackage{gensymb}
\usepackage{hyperref}

\usepackage[percent]{overpic}

\usepackage[utf8]{inputenc}
\usepackage[T1]{fontenc}
\DeclareUnicodeCharacter{2033}{''}

\ifpdf
  \pdfoutput=1\relax                   
  \usepackage{graphicx}                
  \DeclareGraphicsExtensions{.pdf,.png,.jpg,.jpeg} 
\else
  \usepackage{graphicx}                
  \DeclareGraphicsExtensions{.eps}     
\fi%

\graphicspath{{figures/}{pictures/}{images/}{./}} 

\usepackage{microtype}                 
\PassOptionsToPackage{warn}{textcomp}  
\usepackage{textcomp}                  
\usepackage{mathptmx}                  
\usepackage{times}                     
\usepackage{cite}                      
\usepackage{tabu}                      
\usepackage{booktabs}                  

\onlineid{1296}

\vgtccategory{Research}
\vgtcpapertype{evaluation}

\title{Toward Standardized Classification of Foveated Displays}

\author{Josef Spjut \and Ben Boudaoud \and Jonghyun Kim \and Trey Greer \and Rachel Albert \and Michael Stengel \and Kaan Akşit \and David Luebke}
\authorfooter{
\item The authors are with NVIDIA Corporation. E-mail: jspjut@nvidia.com
}

\shortauthortitle{Spjut \MakeLowercase{\textit{et al.}}: Toward Standardized Classification of Foveated Displays}

\abstract{
Emergent in the field of head mounted display design is a desire to leverage the limitations of the human visual system to reduce the computation, communication, and display workload in power and form-factor constrained systems.
Fundamental to this reduced workload is the ability to match display resolution to the acuity of the human visual system, along with a resulting need to follow the gaze of the eye as it moves, a process referred to as \emph{foveation}. 
A display that moves its content along with the eye may be called a \emph{Foveated Display}, though this term is also commonly used to describe displays with non-uniform resolution that attempt to mimic human visual acuity.
We therefore recommend a definition for the term \emph{Foveated Display} that accepts both of these interpretations.
Furthermore, we include a simplified model for human visual \emph{Acuity Distribution Functions} (ADFs) at various levels of visual acuity, across wide fields of view and propose comparison of this ADF with the \emph{Resolution Distribution Function} of a foveated display for evaluation of its resolution at a particular gaze direction.
We also provide a taxonomy to allow the field to meaningfully compare and contrast various aspects of foveated displays in a display and optical technology-agnostic manner.

} 


\CCScatlist{ 
 \CCScat{K.6.1}{Management of Computing and Information Systems}%
{Project and People Management}{Life Cycle};
 \CCScat{K.7.m}{The Computing Profession}{Miscellaneous}{Ethics}
}

\teaser{
  \centering
  \includegraphics[height=1.65in]{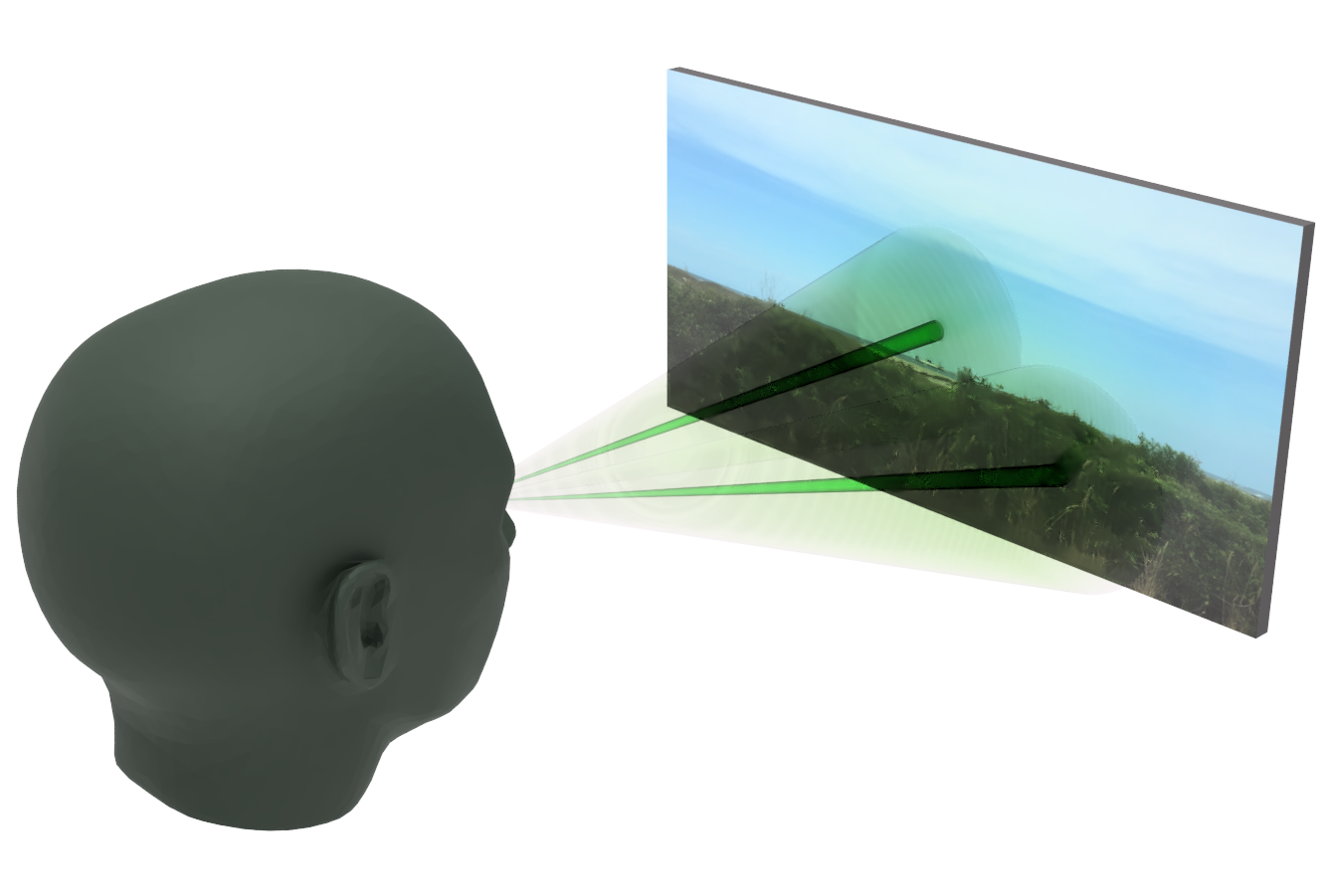}
  \includegraphics[height=1.45in]{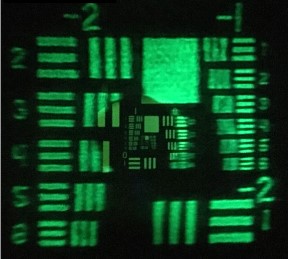}
  \includegraphics[height=1.45in]{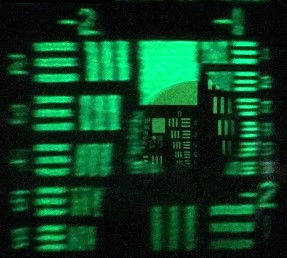}
  \caption{A foveated display is a display designed to function in the context of user gaze. This can mean it follows gaze direction, or expects to be gazed upon in a certain region. Additionally, foveated displays have the potential to vary their resolution (in cycles per degree) using a perception inspired resolution distribution function. The right 2 images demonstrate what a prototype foveated display may look like with 2 different user gaze directions.}
	\label{fig:teaser}
}

\begin{document}


\firstsection{Introduction}

\maketitle

Head Mounted Displays (HMDs) have enjoyed a recent resurgence in popularity, likely due in part to improved display resolution and the availability of lower power electronics for tracking and communication \cite{vieri201818}. However, in order to continue this trend without exceeding power and form-factor requirements in this stringent design space, it is important to avoid over-provisioning hardware resources.
User perception is a vital guide in optimizing system design around functional utility ahead of arbitrary design metrics \cite{parkhurst2002variable}. Based on human perception-inspired design decisions, we set out to describe a taxonomy for classification of such displays in order to establish shared terminology across the field.
The cross-disciplinary nature of this problem space requires expertise from varied backgrounds and often leads to confusion. 
We attempt to describe the essential axes of foveated displays clearly, and clarify common misunderstandings to enable optical engineers, graphics developers and perception scientists to speak a common language.

The rear of the human eye acts as an image sensor in concert with ganglion cells and the brain to generate visual input. The set of sensors on the back of the eyeball is referred to as the \emph{retina}. The central portion of the retina, called the \emph{fovea centralis} (or fovea for short), has the highest density of sensor cells, and provides the highest perceptual visual acuity. Continually, the human \emph{foveates} objects, meaning one or both foveae are angled at a particular object. 

Based on the understanding of the fovea as a high resolution image sensor and the act of foveating as moving the fovea to a particular position, we suggest that the term \emph{foveated display} be used to describe any display system which either steers a display based on the gaze direction (often referred to as gaze-contingent displays \cite{reingold2003gaze}) or varies in actual or perceived resolution proportionate to the acuity of the human visual system. This definition is intended to capture both the essence of foveating as a display (i.e. moving to match user gaze direction) while including the concept of acuity matching (i.e. correct resolution for a given gaze direction) popular in the foveated rendering space~\cite{patney2016towards,weier2017perception}.  We use the term foveated display interchangeably with the traditional term: Gaze-Contingent Multi-Resolution Display (GCMRD) \cite{reingold2003gaze}.

\section{Related Work}
Recent work on foveated rendering~\cite{patney2016towards,guenter2012foveated,albert2017latency,kim2017perceptual,weier2016foveated,stengel2016adaptive} served as a primary inspiration for this classification. Foveated Rendering has come to mean a style of 3D graphics rendering intended to exploit the limited visual acuity of the user across the field of view when assuming a particular gaze direction, and has led to a wide belief that the term ``foveate'' refers solely to variable resolution characteristics. Our classification works to reconcile this modern understanding of the term ``foveate'' with the perceptual science interpretation indicating adaptation to the gaze direction.

A number of previous articles have surveyed and provided loose classification for a variety of types of displays in the past. A number of high quality surveys of Augmented Reality HMDs have been published  \cite{azuma1997survey,cakmakci2006head}. This work often observes that a higher visual acuity is present in the fovea of the user, though none of these surveys attempt to classify different approaches to foveation.

Much prior work has suggested gaze contingent displays, and we review a small subset here.
Reder proposed a gaze-contingent visual stimulus in 1973~\cite{reder1973line} while others~\cite{baldwin1981area,spooner1982trend} suggested desktop displays enabling a foveated display through physical motion. 
Later work~\cite{shenker1987optical,howlett1992high,iwamoto1993development,iwamoto1997development,rolland1998high,akcsit2019manufacturing} applied these concepts to head mounted or near eye display contexts, achieving as much as $24$~cpd of display resolution in the fovea.
Even more recently Godin et al.~\cite{Godin2006High} described a dual projector system with a fixed display foveal inset and Lee et al.~\cite{Lee2017Foveated} applied a similar design using a holographic lens for the near eye context.

To our knowledge, no perceptually motivated, user-facing classification for foveation, as it is described here, has been published or adopted. Instead display-system surveys, such as those cited above, tend to classify designs based on a mix of optical path similarities and individual design performance, often as reported by objective, optical measurements.

\section{Classification Principles}
\label{sec:classification}
We begin our classification based on the assumption that there will be a region of the display called the \emph{foveal inset} (or fovea), which, similar to the fovea of the human eye, achieves a certain level of visual performance. While human visual acuity may continue to increase all the way to the center of the optical system, in practice it makes sense to think of the display foveal inset as a circular conic region around the center of the optical axis of the display of some size. One may describe the foveal inset as a region of (near) constant resolution, intended to match the peak visual acuity of a given user. We then further assume that foveated displays will reduce in resolution as they move away (in eccentricity) from this foveal inset, similarly to human visual acuity.

In order to further describe these assumptions, we propose 2 classification principles as the core to evaluation of all foveated displays: \emph{visual acuity} and \emph{range of gaze}.
\emph{Gaze direction}, as used here, refers to the central ray of a user's instantaneous view, that maps to the center of the fovea.


\subsection{Visual Acuity}
\label{sec:Visual Acuity}
We adopt the standard approach for classifying human vision in describing the user's (ordinary) visual acuity. That is, a nominal human visual acuity, often called letter acuity, would be matched when the foveal resolution matches a 20/20 Snellen fraction (or 6/6 internationally). A user that only achieves half this acuity would be described as 20/40. Note that most readable content is designed to work for people with 20/40 vision, thus 20/40 acuity may be a good goal for foveated display designs.

Equation~\ref{eqn:visiontocpd} provides a simple mapping between Snellen fraction and the maximum \emph{foveal acuity} of a user in cycles per degree ($\mathit{cpd}$). Equation~\ref{eqn:visiontodpi} shows how to convert from this foveal acuity to dots per inch ($\mathit{dpi}$) resolution at a distance of \emph{D} inches.
A common back-of-the envelop conversion from cycles to pixels is to double the number, though more than a simple doubling may be required for displays with arbitrary offsets relative to the virtual content.

\begin{equation}
Resolution_{cpd} = 30\mathit{cpd} \times \mathit{SnellenFraction}
\label{eqn:visiontocpd}
\end{equation}
\begin{equation}
Resolution_{dpi} = \frac{1}{D\times\tan(\frac{1}{2\times\mathit{Resolution_{cpd}})}}
\label{eqn:visiontodpi}
\end{equation}

\subsubsection{Acuity Distribution Functions}
\label{sec:ADF}
In order to completely specify a user's visual acuity it is useful to consider more than just their foveal acuity. If the peripheral acuity requirements are not met, a display may present bothersome artifacts. In order to address  visual acuity in a more holistic manner, the \emph{Acuity Distribution Function} (ADF) is introduced. An ADF describes the angular resolution (in cycles per degree) \emph{as perceived} by a user as a function of \emph{gaze eccentricity}, or angular displacement from the center of gaze fixation (i.e. from the gaze direction).

 If we make a simplifying assumption that we are only interested in the maximum guaranteed (minimum overall) resolution along any radial ``slice'' of gaze eccentricity, and that acuity is strictly decreasing with eccentricity, we can describe an arbitrary ADF as a monotonic decreasing function of the eccentricity of the display/gaze. While this model does not correctly match the human visual system which varies in acuity based on which radial angle is selected, and includes things like the blind spot, it does give us an effective way to classify displays by approximating the ADF of the user. One such approximation for an average user's ADF over eccentricity($e$) is the following function:

\begin{equation}
ADF(e) = 
\begin{cases}
    F & e \leq e_0 \\
    \dfrac{S}{e-e_0+\dfrac{S}{F}} & e > e_0
\end{cases}
\end{equation}

\begin{figure}
\includegraphics[width=\columnwidth]{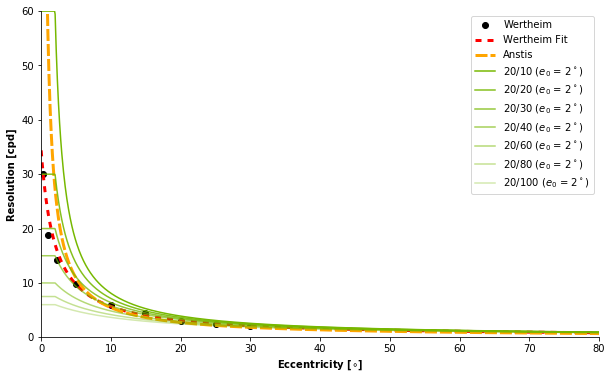}
\caption{ADF Approximation over a range of common visual acuity plotted with historically measured data/fits for 20/20 acuity ($e_0=2^\circ$, S = 75 cpd/$^\circ$)}
\label{fig:ResEcc}
\end{figure}

Where $\mathit{F}$ refers to the foveal acuity (as calculated in Equation \ref{eqn:visiontocpd}), $\mathit{S}$ refers to the peripheral roll-off ``slope'' in cpd/$^\circ$, and $\mathit{e_0}$ refers to the width of the foveal region (assumed to be constant resolution). Analysis of historical data \cite{wertheim1894uber, anstis1974chart} was used to determine a reasonable peripheral roll-off slope of 30 cpd/0.4 $^\circ$ or about 75 cpd/$^\circ$. The resulting ADF is visualized for $e_0=2^\circ$ over the range of common visual acuity, together with historical acuity measures from Wertheim \cite{wertheim1894uber} and Anstis \cite{anstis1974chart} in Figure~\ref{fig:ResEcc}. Note that while our suggested ADF is inspired by biological~\cite{curcio1990topography} and perceptual~\cite{anstis1974chart,wertheim1894uber} measurements of human visual acuity, it is intentionally an approximation loosely based on this prior work. Others~\cite{weier2017perception,meng2018kernel,reddy2001perceptually} have suggested similar approaches, and more in-depth work has been conducted based on physiological cues such as photo-receptor density in the retina and cortical magnification ~\cite{rovamo1979estimation}.

The ADF described above is somewhat pessimistic (i.e. tends to predict acuity strictly above measured estimates). For a more tightly fit ADF model refer to Section \ref{sec:Alt ADF}.

\subsubsection{Resolution Distribution Functions}
\label{sec:RDF}
A \emph{Resolution Distribution Function} (RDF) describes the full spatial or angular resolution presented by a display as a function of \emph{display eccentricity}, or angular displacement from the central optical axis of the display. Just as an ADF categorizes user visual acuity, an RDF describes corresponding display resolution. 

In reality, RDFs are 3 or 4 dimensional functions translating an azimuth/elevation of display/gaze eccentricity (and possibly focal plane) into a resolution in cycles per degree. However, by applying the same simplification principles described for ADFs above, RDFs can easily be compared with ADFs using a 2-dimensional plot. A display is said to be \emph{acuity matched} to a particular level of human visual acuity (i.e. 20/20 acuity matched) when its RDF exceeds a specified ADF over the entire display area. Failing this, an RDF/ADF comparison allows us to evaluate things like where in the field of view a display is under-performing. Note that regions where the RDF exceeds the required ADF represent ``wasted'' pixels, for more information on pixel waste refer to Section \ref{sec:Pixel Waste and RDF Efficiency}.

\subsection{Gaze Direction and Range of Gaze}
When discussing display systems, many simplify the gaze model to identify a single 2D point in a plane/display, representing the direction within the view frustum of the virtual image presented on the display. For our purposes, we assume the gaze direction is represented as an azimuth and elevation of rotation of the eye (on a per eye basis), or can be converted into this format.

Historically, displays that respond to changes in gaze direction have been referred to as \emph{gaze-contingent displays}.
When a display is designed with a non-uniform RDF intended to be viewed across multiple gaze directions, it is useful to define the \emph{range of gaze} as the maximum extents of gaze direction that support a particular perceived RDF, either through physical steering or software rendering.
It is generally assumed that a display supports \emph{all} gaze directions within its range of gaze, implying these maximum extents as bounds for the user's foveation. If this is not the case, supported gaze directions should be explicitly reported.

\section{Taxonomy}
\label{sec:taxonomy}
In order to help compare and evaluate foveated displays we suggest a hierarchical, multi-class framework. This framework seeks to be:

\vspace{-0.4em}
\begin{itemize}
\itemsep-0.4em
\item Easy to understand and useful in categorizing various displays
\item Quickly evaluable by a user without knowledge of display design or access to optical measurement equipment
\item Robust to various display content the user does not control
\item Useful for a designer at any stage of development without a need for large-scale user studies
\end{itemize}
\vspace{-0.4em}

We attempt to capture both display RDF matching and gaze contigence as part of a single combined classifier described in the following section.

Pursuant to the objectives above, this framework inherently incorporates the bias of an observer into its decision process. For example, a user with 20/40 vision may reach different conclusions about whether a display is acuity matched than one with 20/10 visual acuity. Fortunately many of these biases can be quickly categorized via observer self-reporting and observation bias can be reduced via majority vote among larger sets of (similar) observers. For more information refer to section \ref{sec:RangeOfAcuity}.

\subsection{Resolution Contingent Classification}
\begin{figure}
\begin{overpic}[width=\columnwidth]{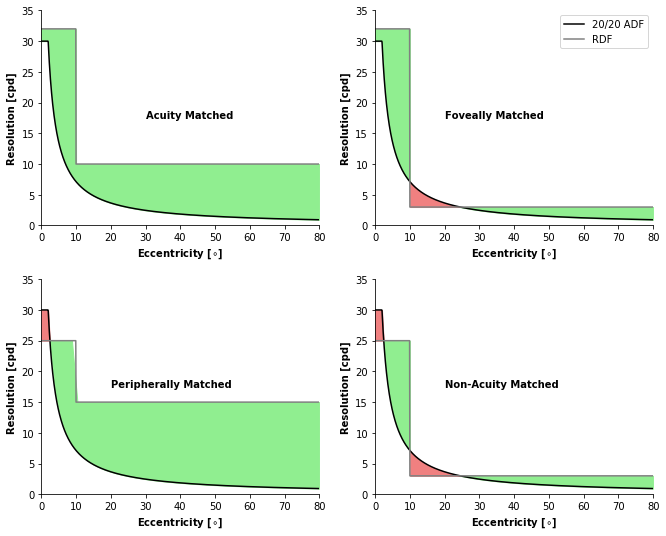}
\put (24,70) {\small Class A}
\put (24,30) {\small Class C}
\put (70,70) {\small Class B}
\put (70,30) {\small Class D}
\end{overpic}
\caption{Four possible comparisons of user ADF and display RDF}
\label{fig:ADFvsRDF}
\end{figure}

In this section we introduce 4 classes (A-D) of acuity matching intended to allow quick classification of  the RDF of a foveated display.
Figure~\ref{fig:ADFvsRDF} gives a pictorial representation of how a 2-tiered foveated display might be placed into these classifications given a static fixation point (0 degrees eccentricity). Note that the choice of ordering between classes B and C is arbitrary and that we only intend to distinguish between them rather than imply that one is strictly superior to the other. Class A is intended to be strictly better than the others and Class D is intended to be strictly worse.

\subsubsection{Class A: Acuity Matched}
A Class A or \emph{acuity matched} display is one in which the maximum resolution presented by the display meets or exceeds the foveal acuity of user, while the periphery presents no noticeable artifacts. 
Note that though our tendency may be to refer to this case as ``indistinguishable from reality'' in practice acuity matched displays are limited to the effects of resolution, whereas perfect emulation of reality requires additional optical effects (i.e. defocus).

\subsubsection{Class B: Foveally Matched}
A Class B or \emph{foveally matched} display is one in which the foveal region of the display meets or exceeds the user's visual acuity while the peripheral region presents noticeable artifacts. This is characterized by being able to read a Snellen eye chart at the user's normal visual acuity, but noticing artifacts when viewing items in the near/far periphery. Examples of peripheral artifacts to look for include:

\begin{itemize}
\itemsep-0.4em
\item Incorrect acuity roll-off in the periphery (i.e. tunnel vision)
\item Resolution changes/blending between fovea and periphery
\item Color differences between the fovea and periphery
\item Temporal artifacts caused by motion (i.e. flicker)
\end{itemize}

\subsubsection{Class C: Peripherally Matched}
A Class C or \emph{peripherally matched} display is one in which the peripheral region of the display presents no noticeable artifacts, but the foveal region fails to meet or exceed the user's visual acuity. This is characterized by being unable to read a Snellen eye chart at the user's normal visual acuity, while not noticing any significant artifacts in the periphery.

\subsubsection{Class D: Non-Acuity Matched}
A Class D or \emph{non-acuity matched} display is one in which the foveal inset fails to meet the acuity of the user and peripheral artifacts are noticeable.

\subsection{Gaze Contingent Classification}
In order to help address how a system \emph{foveates}, that is to say whether and how it adapts to changes in user gaze direction, we introduce an additional 4 class taxonomy using the numbers 1-4.

\subsubsection{Class 1: Fully Foveated Displays}
A Class 1 or \emph{fully foveated} display is one where the RDF does not \emph{noticeably} change for any gaze direction within the (full) range of gaze of the user. It is worth noting that ``noticeably'' deliberately implies that the RDF may in fact change, but these changes must exceed the acuity of the user's ADF across the full range of gaze.

The simplest example of a Class 1 display is any sufficiently sized, pixel-based, fixed resolution display that subtends the user's full field of view. Since the pixel pitch of such a display does not change with user gaze the display presents the same (uniform) RDF for all possible gaze directions. For more information on this case refer to Section \ref{sec:Brute Force Foveation}.

\subsubsection{Class 2: Practically Foveated Displays}
A Class 2 or \emph{practically foveated} display is one in which the RDF remains constant for a large enough sub-set of the user's range of gaze that the gaze direction is not likely to exceed the display's supported range of gaze under normal use across a broad set of applications.
We suggest that exceeding a $\pm$15$\degree$ range of gaze directions may be sufficient to qualify as practically foveated \cite{hatada1980psychophysical,sprague2015stereopsis}, as maintaining a gaze direction beyond these extents is uncomfortable to the human for extended periods of time.

\begin{table*}[ht!]
\caption{Combining RDF classification (letters) with motion classification (numbers) produces this classification matrix}
\label{tab:classification}
\begin{tabular}{|m{2cm}|>{\raggedright\arraybackslash}m{3.4cm}|>{\raggedright\arraybackslash}m{3.4cm}|>{\raggedright\arraybackslash}m{3.4cm}|>{\raggedright\arraybackslash}m{3.4cm}|}
\cline{2-5}

\multicolumn{1}{c|}{}
& \thead{Class A\\Acuity Matched} 
& \thead{Class B\\Foveally Matched} 
&  \thead{Class C\\Peripherally Matched} 
&  \thead{Class D\\Non-Acuity Matched} \\
\hline
\thead{Class 1\\Fully\\Foveated} & 
For any gaze direction, the display meets or exceeds the user's visual acuity without any peripheral artifacts
& For any gaze direction the foveal inset matches user acuity, but peripheral artifacts are present
& The foveal inset fails to match user acuity, but achieves equal resolution over all gaze directions with no peripheral artifacts
& Neither the foveal inset nor periphery matches user acuity, but the display achieves equal resolution over all gaze directions\\
\hline
\thead{Class 2\\Practically\\Foveated} & 
For a practical sub-set of gaze directions the display meets or exceeds the user's visual acuity without any peripheral artifacts
& For a practical sub-set of gaze directions the foveal inset matches user acuity w/ peripheral artifacts present 
& The foveal isnet fails to match user acuity, but achieves equal resolution over a practical sub-set of gaze directions with no peripheral artifacts
& Neither the foveal inset nor periphery matches user acuity, but the display achieves equal resolution over a practical sub-set of gaze directions\\
\hline

\thead{Class 3\\Partially\\Foveated}
& For a small sub-set of gaze directions the display meets or exceeds the user's visual acuity without any peripheral artifacts
& For a small sub-set of gaze directions the foveal inset matches user acuity w/ peripheral artifacts present
& The foveal inset fails to match user acuity, but achieves equal resolution over a small sub-set of gaze directions with no peripheral artifacts present
& Neither the foveal inset nor periphery matches user acuity, but the display achieves equal resolution over a small sub-set of gaze directions\\
\hline

\thead{Class 4\\Non-Foveated}
& For a single gaze direction the display meets or exceeds the user's visual acuity without any peripheral artifacts
& For a single gaze direction the foveal inset matches user acuity w/ peripheral artifacts present
& The foveal inset fails to match user acuity and foveal acuity changes with gaze, but no peripheral artifacts are ever present
& Neither the foveal inset nor periphery matches user acuity, and the RDF appears to change for any given gaze direction\\
\hline
\end{tabular}
\end{table*}

If the large display from the example above was to be reduced in size to the point where any (comfortable) user gaze direction fell within its area, but a user could push their eyes to an extreme angle (say {25\degree} from center) and view a region outside the display it would be considered practically foveated.

\subsubsection{Class 3: Partially Foveated}
A Class 3 or \emph{partially foveated} display is similar to a Class 2 display, but wherein the RDF remains constant for a sub-set of gaze directions smaller than the $\pm$15$\degree$ practical range of gaze.
While the RDF changes noticeably beyond the more limited range of gaze for Class 3, many believe this to be a useful class for particular applications. The range of applications to which this class could be applied expands under the belief that the user's behavior should change to improve the display's experience. This class of display may cause changes to the way the user balances head and eye motion. These changes may be acceptable or problematic, we suggest that they be thoroughly studied for any adverse effects before becoming widely adopted.

If the monitor used in the previous two examples were reduced to the size of a cell phone held at arms length, it would no longer cover all practical gaze directions a user could achieve. In this case the display has become partially foveated.

\subsubsection{Class 4: Non-Foveated Displays}
A Class 4 or \emph{non-foveated} display, is one in which the RDF changes as the user gaze directions changes. In Class 4 displays, there is no region of the display over which the RDF stays perceivably constant.
Alternatively, the user can tell that the RDF remains in place spatially even as the gaze direction shifts.

An example of a non-foveated display is a modern HMD (such as a an HTC Vive) in which the (short focal length) lens used to place the virtual image plane of the display at a comfortable viewing distance also imparts the (undesired) side-effects of aberration and astigmatism. This occurs when the user's viewpoint is not axially aligned with the lens/display or the user's pupil travels within the eye box presented by the display. The resulting impacts can be interpreted as an RDF that always changes with gaze direction.

\subsection{Combined Classification}
In order to quickly summarize both the foveation and acuity matching of a display a combined classifier is proposed. A display can be quickly summarized by appending its resolution contingent classification (letter) to its gaze contingent classification (number). For example a class A1 foveated display is a display that matches visual acuity consistently for every possible eye angle.
Table~\ref{tab:classification} summarizes the space of possible classifications and gives a more detailed descriptions for each.



\subsection{Classification Procedure and Examples}
Classification of gaze contingency and resolution distribution are considered independently evaluated within this taxonomy.

\subsubsection{Classifying Resolution Distribution}
Resolution contingent classification can be evaluated (in the fovea) by viewing a Snellen eye chart and comparing the resulting acuity determination with the known (assumed pre-measured) visual acuity of the user. Peripheral evaluation can be conducted by moving any object (of known geometry) from the fovea into the periphery while carefully observing for any artifacts that may become present.

A small subset of displays, for example some pupil forming architectures or retinal projection displays, may be able to allow certain users to read text \emph{beyond} their typical, unaided visual acuity. In these cases a user could theoretically evaluate a display above their own visual acuity, but this is not considered a likely or relevant case for the time being.

\subsubsection{Classifying Gaze Contingence}
Gaze contingent classification can be evaluated by looking for changes in apparent resolution of the display while sampling different gaze directions throughout the user's range of gaze. It may be useful to provide some constant content at various positions in the display's field of view while evaluating this. The suggested procedure for classifying gaze contingency is as follows:

\begin{enumerate}
    \itemsep-0.4em
    \item Find the gaze direction at which the display provides the highest perceived foveal resolution.
    \item Gaze away from this spot by a small amount ($<$5\textdegree) in all directions, if the RDF changes for any gaze direction within this range the display is Class 4, if not continue.
    \item Continue to gaze to a peak comfortable working angle (think the furthest angular at which it would be comfortable to work on a monitor without head motion), if the RDF changes from the previous step the display is Class 3, otherwise continue.
    \item Move to the maximum range of gaze (not necessarily sustained), if the RDF changes from the previous step the display is Class 2, otherwise the RDF has not changed and the display is Class 1.
\end{enumerate}

\subsubsection{Classifying Existing Designs}
Currently available VR displays such as Vive Pro, Valve Index and Oculus Rift S provide sufficient peripheral resolution to prevent noticeable artifacts, but their foveal resolution does not match the visual acuity of most users. This results in a 20/20 Class C4 classification. To help match foveal acuity, Varjo applies a high resolution foveal inset to a VR display \cite{konttori2017display}. Their early prototype (without gaze tracking) matched foveal and peripheral acuity, but made no attempt to steer the inset with gaze direction so it can be classified as Class 20/20 A3 since the foveal inset is large enough to support a small range of gaze. 
Table \ref{tab:ExistingDesignClassification} summarizes some additional classification information at the 20/20 acuity level. 

\begin{table}[t]
\caption{Classification of Existing HMD Designs}
\label{tab:ExistingDesignClassification}
\begin{tabular}{|l|c|c|c|c|}
\hline
\thead{Design} & \thead{FoV [\textdegree]} & \thead{Res. [cpd]} & \thead{Steerable} & \thead{Class}\\
\hline
Vive & 100 & 5.4 & No & D4\\
Vive Pro & 100 & 7.2 & No & C4\\
Hololens & 30 & 21.2 & No & D4\\
Varjo VR-1 & 32/100 & 30/7.2 & No & A3\\
Kim \cite{kimjeong19} & 30/86 & 30-60/3 & Yes & B2\\
\hline
\end{tabular}
\end{table}

\section{Practical Considerations}
While our taxonomy as described in section \ref{sec:taxonomy} is usable on its own, there are a variety of practical considerations and related topics that warrant additional discussion. Some of these are described below.

\subsection{Range of Acuity}
\label{sec:RangeOfAcuity}
The classification principles, particularly the resolution classification, applied in Table~\ref{tab:classification} can be degraded for any given user's visual acuity. For example a display may be considered Class A1 for a 20/80 user though Class D1 for a 20/20 user. This limitation is inherent to the goal of taxonomy evaluation from a wide variety of user perspectives, and makes reporting of the acuity level at which an evaluation was performed essential to the classification result. 

However, in cases of extreme disparity of evaluation acuity, the impacts of differences in acuity may go beyond resolution. For example, a user with 20/20 vision would be likely to classify many modern VR designs as Class C4 or D4; however, a user with 20/200 vision may be tempted to classify the same design as Class A2-A3. For this reason, it may be practical to limit the range of acuity to consider valid for evaluation. 20/10 to 20/40 vision can be considered practical ranges for designs intended for mass markets.

\subsection{Tiered vs Continuous Foveation}
One area of interest for foveated display designers, but not necessarily relevant to the taxonomy presented in this work, is the use of multiple, discrete tiers of display resolution versus attempting to approximate the user's ADF throughout the display field of view.

An \emph{N-tiered} foveated display attempts to combine N discrete levels of resolution into a single effective display area. For example, a simple foveated display design might be a 2-tiered foveated display optically combined to overlay one on the other (as demonstrated in Figure \ref{fig:ADFvsRDF}.

A \emph{continuously foveated display} is a non-tiered (or very finely tiered) display which attempts to smoothly approximate the target acuity across all points of the field of view. Though a continuously foveated display represents the limiting case for acuity matching with minimal pixel waste, it is somewhat impractical to consider when using pixel-based display technologies as they imply a discrete sampling process. For the sake of more practical discussion it is fair to say that a sampled foveated display that exceeds the Nyquist criteria for human visual acuity across its full field of view is continuously foveated. An example of a continuously foveated display would be a uniform resolution display viewed through a set of optics imparting pincushion distortion across the full field of view.

Interestingly, Hoffman et al. reported that in the non-foveated, acuity matched situation (class A4), a 2-tiered foveated display is less noticeable than an N-tiered foveated display if the transition occurs in the periphery region \cite{hoffman2018limits}. Meanwhile, in practically foveated (Class 2) designs with gaze tracking \cite{sun2017perceptually, patney2016towards}, N-tiered display provided the same level of perceptual quality to the uniform resolution display. It has not been clearly analyzed which transition method is the best in general. However, it is clear that the blending algorithm for a certain foveated display should be chosen in consideration of the device's foveation error, temporal stability and target user ADF.

\begin{figure}
\includegraphics[width=\columnwidth]{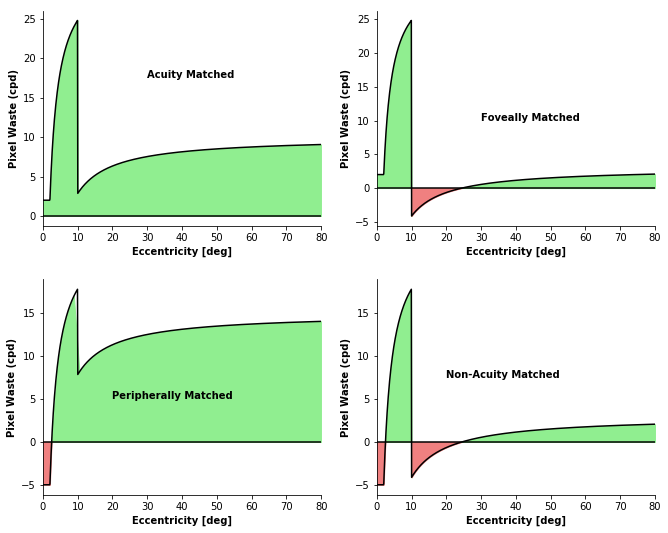}
\caption{Pixel waste of the theoretical 2-tiered designs in Figure \ref{fig:ADFvsRDF}}
\label{fig:PixelWaste}
\end{figure}

\subsection{Pixel Deficit, Waste, and RDF Efficiency}
\label{sec:Pixel Waste and RDF Efficiency}
We introduce \emph{pixel deficit} as the number of cycles (of resolution) in a display that fail to meet the target ADF (user's visual acuity):

\begin{equation}
deficit_{px} = \int |min(RDF(e)-ADF(e),0)|de
\end{equation}

Pixel deficit describes the total under-specification of pixels to meet the target ADF across (a portion of) the field of view of the display. It can be visualized as the red area(s) in Figure \ref{fig:PixelWaste}. A pixel deficit of 0 over the fovea/periphery means a display is foveally/peripherally (class B/C) matched. A total pixel waste of 0 means a display is acuity matched (Class A).

Alternatively, a ``brute force'' display which presents a uniform, foveal resolution image across its full field of view in order to meet a user's foveal acuity may be considered ``wasteful'' from a pixel budget perspective. This is because the user's ADF is not able to observe these extra pixels beyond a small portion of their field of view. Along with the desire to reduce pixel count based upon the limits of human visual acuity comes an aspiration to characterize both this waste and the resulting efficiency a display achieves. For this reason, let us define \emph{pixel waste} as:

\begin{equation}
waste_{px} = \int max(RDF(e)-ADF(e),0)de
\end{equation}

The pixel waste represents the effective over-provisioning of resolution within a given display area in cycles or pixels. It can be visualized as the area between the target ADF and measured RDF of a given display. Areas of the field of view in which ADF exceeds RDF result in positive pixel waste (implying resolution was in fact ``wasted'') while areas where RDF exceeds ADF result in 0 pixel waste (implying insufficient resolution was provided). The total evaluation of pixel waste can be visualized as the total light green area presented in in Figure \ref{fig:PixelWaste}.

By separating pixel waste from pixel deficit we are better able to describe the two common types of ADF vs RDF mismatch. While only the pixel deficit is useful in providing the resolution-contingent classification of a display, the pixel waste is a useful way to evaluate the efficacy of a display in matching the user's acuity.

In order to help make evaluation of pixel waste more relative to display resolution, we propose the use of a pixel utilization or \emph{RDF efficiency} metric to help quantify and compare designs efficacy in the distribution of pixels. This efficiency can be calculated as shown below.

\begin{equation}
\epsilon_{RDF} = 1-\int_{display}\frac{max(RDF(e) - ADF(e),0)}{RDF(e)} de = 1 - \frac{waste_{px}}{count_{px}}
\end{equation}

Where $count_{px}$ is the total (1D) cycle count of the display. This means that the RDF efficiency is 1 (or 100\%) when ADF matches RDF. Otherwise RDF efficiency is less than 1 and represents the average portion of pixels in the display for which resolution is sufficient (i.e. pixels are not wasted), the remainder representing pixel waste.

It is worth noting that neither pixel waste nor RDF efficiency can be used to perform resolution contingent classification. This is because the pixel waste only represents regions where the RDF exceeds the ADF. Pixel waste and RDF efficiency are instead proposed to help designers become more aware of where a given display is wasting bandwidth, energy, and/or time by provisioning resolution beyond what is perceivable by the user.


\begin{figure}
    \includegraphics[width=\columnwidth]{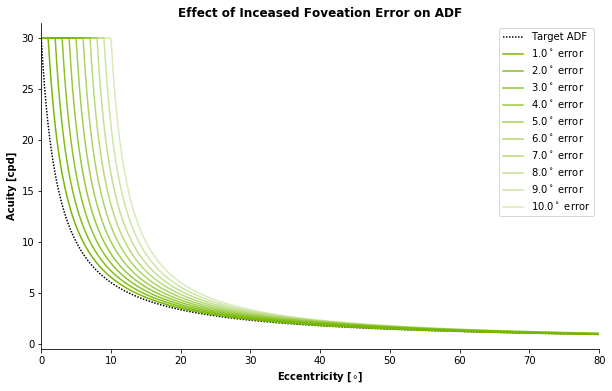}
    \caption{Target ADF and ideal RDF degraded for $1-10 ^{\circ}$ foveation error}    
    \label{fig:FovError}
\end{figure}

\subsection{Foveation Error}
Let us define \emph{foveation error} as the maximum angular distance between the central optical axis of the eye and the central optical axis of the display, particularly the region of foveal acuity. Foveation error can be thought of as the instantaneous difference between the ideal display position and actual display position given a user's gaze direction.

Foveation error can arise from a number of sources including:
\vspace{-0.4em}
\begin{itemize}
\itemsep-0.4em
\item Error or latency in reported user gaze direction
\item Error, speed, or accuracy limitations of steering mechanisms
\item Display render and update latency (e.g. LCD)
\end{itemize}
\vspace{-0.4em}

There are a variety of ways that one might measure the error of foveation for a display. Some possibilities include visual angle (degrees or radians), latency (seconds), or distance in display space (pixels). While each of these and other measurements could be useful for particular purposes, we suggest that visual angle is the most useful for classifying a foveated display.

Given a display with a fixed size foveal region, the foveation error can be treated as an uncertainty region that needs to be removed from the total foveal field of view before reporting the achieved foveal region size. Alternatively, given a target foveal region field of view, this region needs to be increased in size by the foveation error to be able to guarantee the required foveal field of view. Figure~\ref{fig:FovError} shows how much of an impact $1-10\degree$ foveation error has on the ADF.
Other foveation models exist and related work~\cite{guenter2012foveated} provides additional information.

\subsection{Alternate ADF Models}
\label{sec:Alt ADF}
The ADF model proposed in Section \ref{sec:Visual Acuity} is somewhat pessimistic from a design target perspective (i.e. tends to over approximate human visual acuity). We refer to this model as the ``constant fovea size'' model as it defines a size for the foveal inset and computes the correct foveal intersection point for the curved part of the ADF given the roll-off slope and foveal acuity. While the 20/20 model agrees quite well with historical data, the lower acuity models (20/30, 20/40, ...) tend towards the 20/20 acuity model at large eccentricities, potentially over approximating human visual acuity. More perceptual study data demonstrating the relationship between acuity and visual eccentricity across varying subject Snellen ratio and wide(r) field of view would help substantiate this claim.

To reduce this effect we introduce an alternate \textit{slope} based ADF model. The concept being that we use a (constant) slope similar to the MAR figure described by Anstis \cite{anstis1974chart} to describe the roll-off of the acuity for all different Snellen fractions. This model can be thought of as setting a constant roll-off and computing the correct intersect to achieve this roll off.

\begin{equation}
ADF(e)=
\begin{cases}
    F & e \leq e_0 \\
    \dfrac{1}{\frac{1}{S}(e-e_0)+\frac{1}{F}}=\dfrac{F}{S'(e-e_0)+1} & e > e_0
\end{cases}
\end{equation}

Where $S' = F/S$ and has units of cpd/(degree eccentricity/degrees per cycle) or just 1/degree eccentricity. Here we can use the Wertheim best fit (0.44) or Anstis suggested (0.55) slope and directly fit an ADF using this slope.

\begin{figure}
    \centering
    \includegraphics[width=\columnwidth]{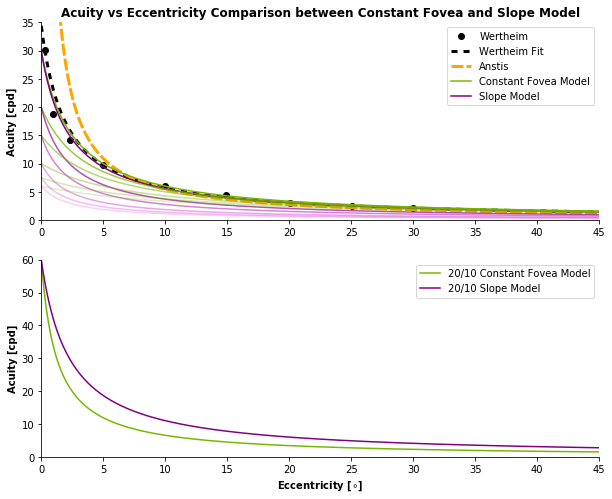}
    \caption{Comparison of ``constant fovea size'' model from Section \ref{sec:ADF} and slope-based model proposed above with 0$^\circ$ fovea size across a range of acuities.}
    \label{fig:SlopeModelComparison}
\end{figure}

We compare the slope model proposed above to our constant fovea size model from Section \ref{sec:ADF} in Figure \ref{fig:SlopeModelComparison}. It can be observed that the slope model does produce consistent ``roll-off'' slope across differing visual acuity and degrades much more quickly than the constant fovea size model for visual acuity lower than 20/20. In the 20/10 acuity case this trend reverses and the slope model exceeds the constant fovea size model ADF. This is expected as the roll-off for the slope model is fixed and cannot converge as quickly as that used for the constant fovea size model.

\section{Discussion}
The following section contains some additional discussion topics that are not as immediately relevant to the taxonomy proposed, but present interesting cases for future study and may potentially become relevant for expanded classifications. 

\subsection{Display Steering}
Although modern HMDs can present reasonable quality virtual images at a comfortable viewing distance when the user's gaze direction is aligned with the optical center of their lens systems, the astigmatic effects introduced by viewing the display from more extreme gaze directions can severely limit the effective resolution.
The solution for most designers aspiring to build foveated displays that are capable of delivering highly adaptive RDF response to gaze direction is display steering. 
\emph{Display steering} is, in the broadest sense the concept of translating some portion of a display system to produce an effective change in RDF.
We define two broad methods for steering these displays as follows:

\emph{Soft Steering} refers to any steering mechanism that does not require (physically) moving parts to produce a displacement/rotation in light entering the eye of the user. Examples include using an over-sized display in which an active sub-set of the display/light source maps to a particular eye position, or the use of electro-optical devices such as Spatial Light Modulators (SLMs) to mask/steer light without a need for moving parts.

\emph{Hard Steering} is defined as any light steering mechanism that makes use of mechanical components. Examples include (tilt-tip) mirrors, gimbal or translation stages, and other dynamic optical components such as liquid lenses. Designers could use one or both of these approaches to achieve the perceived motion of display resolution.

\subsection{Impacts of Blending Regions}
In N-tiered foveated displays with noticeable resolution transitions some portion of the (higher resolution) inset display(s) needs to be given up to blending resolution down to the (lower resolution) periphery. This blend region can be treated as a part of the foveation error (it is effectively removed from the un-blended foveal region size).

\begin{figure}
    \centering
    \includegraphics[width=\columnwidth]{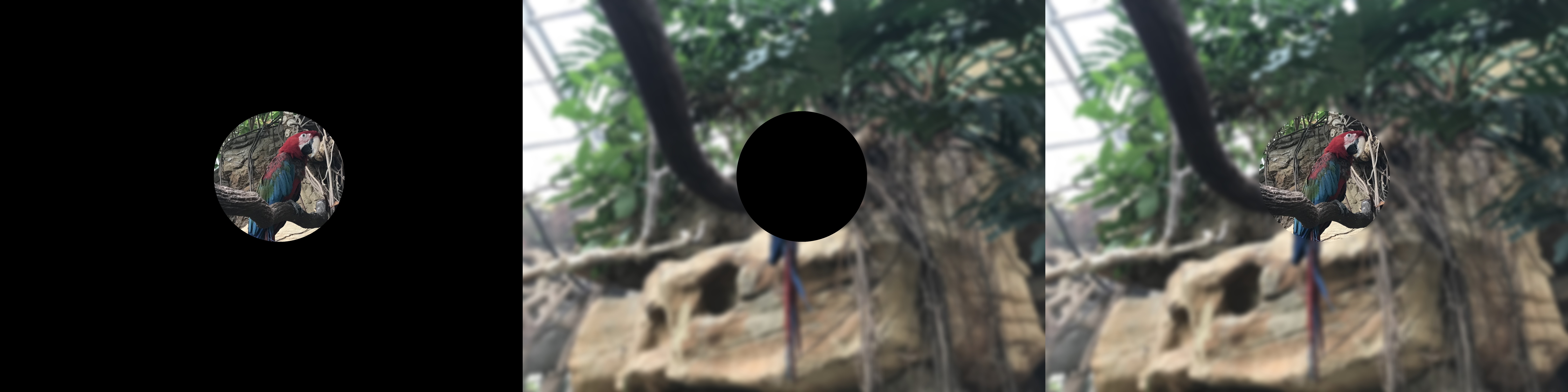}
    \includegraphics[width=\columnwidth]{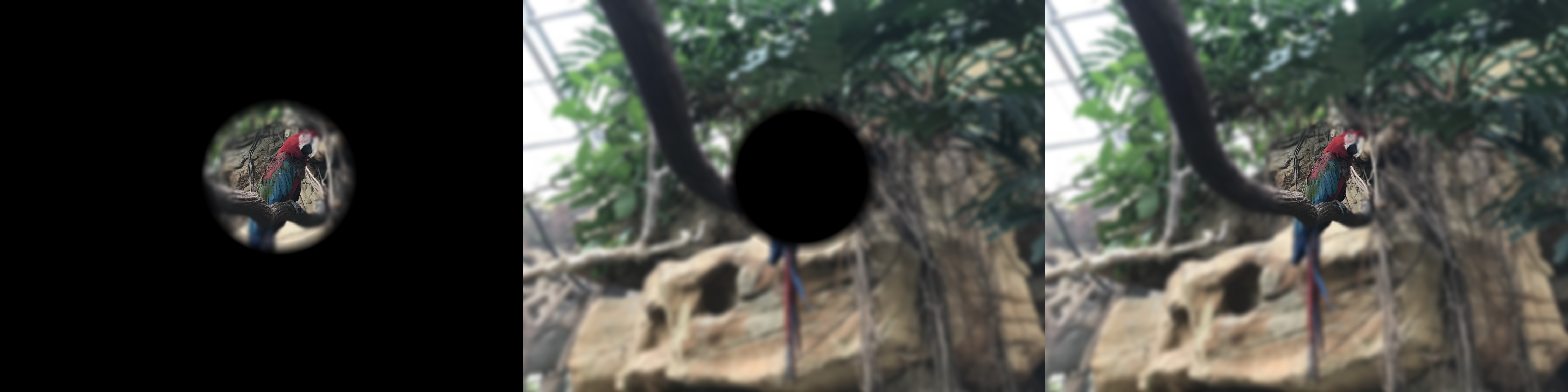}
        \centering
        \begin{tabular}{p{0.28\columnwidth}|p{0.28\columnwidth}|p{0.28\columnwidth}}
            inset & periphery & combined \\
        \end{tabular}
    \caption{Without blending region (top) the inset and peripheral display have a perceivable edge. Adding a blending region (bottom) hides the display boundary.}
    \label{fig:blending}
\end{figure}

The optimal size of blend region can be thought of as proportional to the difference in resolution between the two regions of the display it blends. More accurately, the blend region should be sized so that the RDF of the display (including the blend region) approximates the ADF target for the display as closely as possible. An example of the hard edges that are present without including a blending region and the improvement possible when a blending region is included can be seen in Figure \ref{fig:blending}. Extensions to this taxonomy to account for blending regions more explicitly are left to future work.

\subsection{Color Matching}
We have presented this taxonomy while ignoring many practical considerations in an effort to keep the taxonomy relatively independent of particular design constraints and focused on the user experience.
However, when multiple displays or display technologies, particularly those utilizing independent light sources, are used to create a foveated display (LCOS for foveal inset and DLP/DMD for periphery for example), it is necessary to perform a calibration step to bring the color gamut and contrast of the differing displays into perceptual agreement. 

\begin{figure}
    \centering
    \includegraphics[width=0.49\columnwidth]{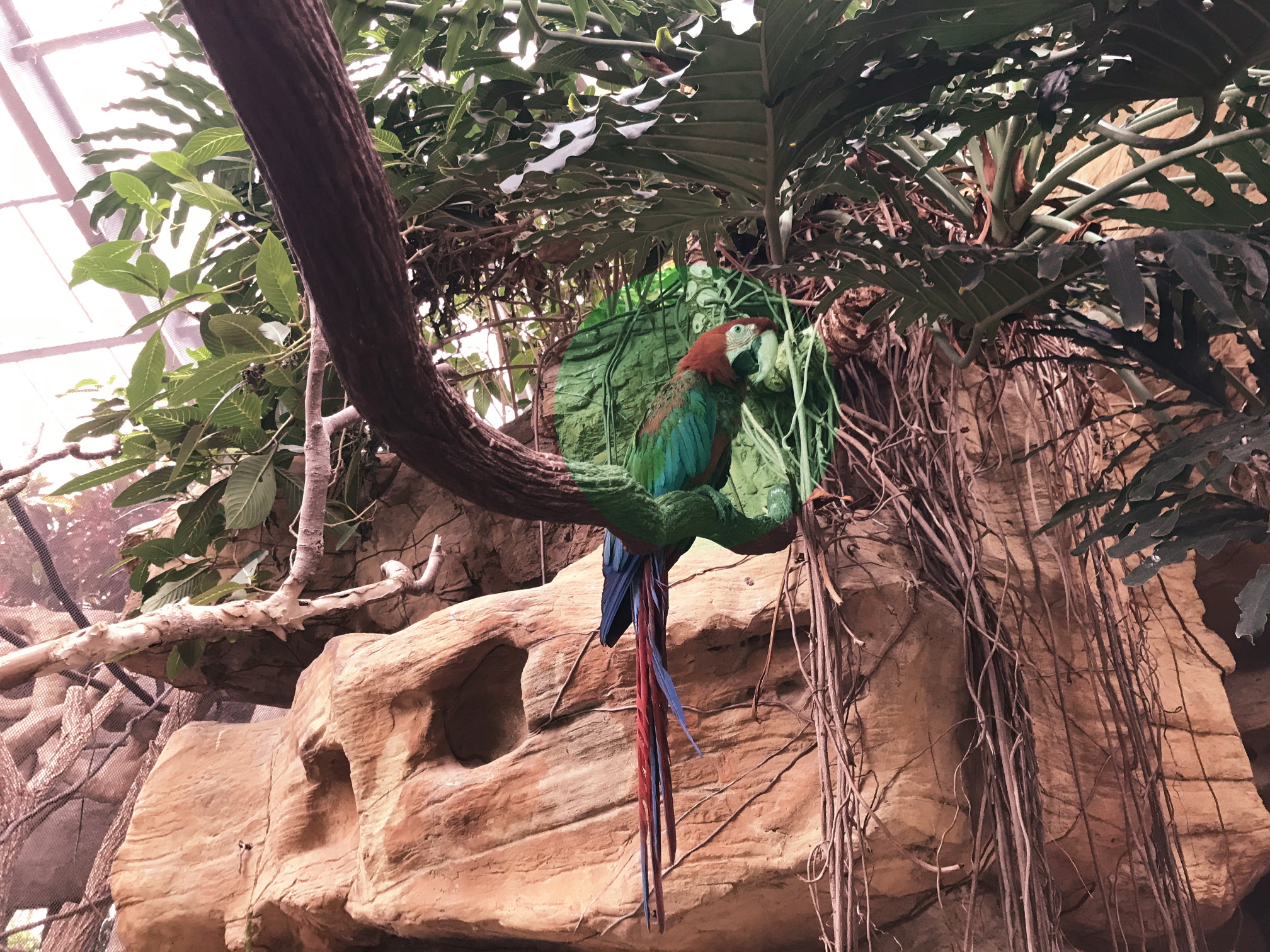}
    \includegraphics[width=0.49\columnwidth]{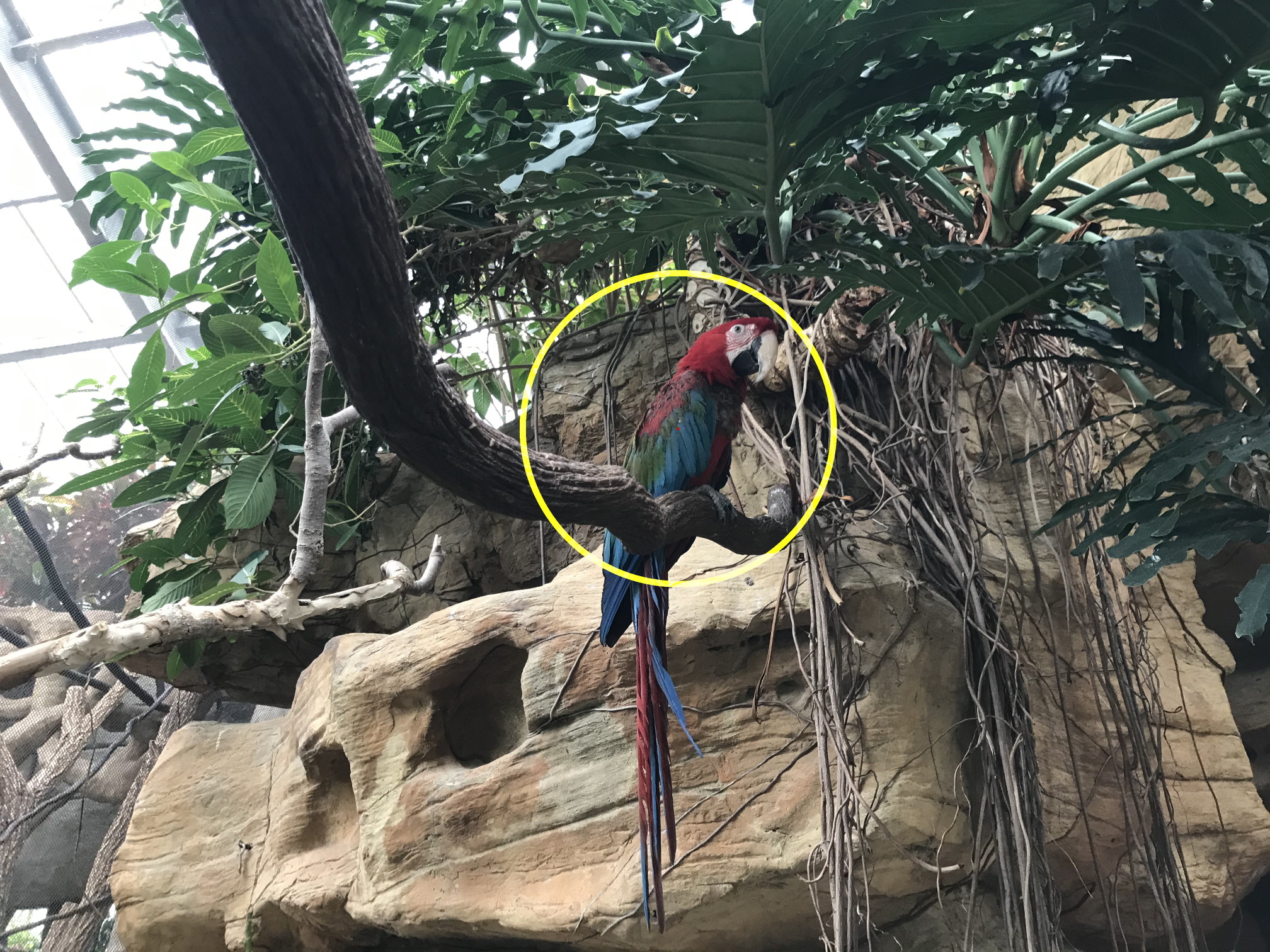}
        \begin{tabular}{p{0.42\columnwidth}|p{0.42\columnwidth}}
            before color matching & idealized color match \\
        \end{tabular}
    \caption{Idealized impact of color matching for a foveated AR display. Before color matching (left) the displays disagree in color and after (right) the color is consistent. The yellow circle identifies the foveal inset. Color matching issues such as those seen on the left can be considered peripheral artifacts. }
    \label{fig:ColorMatching}
\end{figure}

Techniques for performing this calibration are well known \cite{sharma2002lcds}, but cannot always be accomplished due to particular display technologies' color gamut limitations.
Since color matching often enforces a reduced color gamut for the resultant display, impacts on user experience should be carefully considered when selecting displays for a given design. Figure~\ref{fig:ColorMatching} provides a theoretical example of what an idealized color matching process would achieve.

\subsection{Brute Force Solution and Limitations}
\label{sec:Brute Force Foveation}
The simplest strategy for creating a Class A acuity matched display is that of ``brute force'' acuity matching, or displays in which the maximum desired resolution is achieved uniformly across the full field of view (or at least the range of gaze). While this sort of display can achieve the criteria required for full acuity matching, it does so by giving up many of the efficiency and form-factor benefits that foveation sought to provide.

The main challenge in brute force foveated displays is that of supporting a wide range of gaze and/or field of view. If a uniform resolution display is to support the full human range of gaze, it needs a very large pixel budget, and as a result produces very large pixel waste/low RDF efficiency. This results in higher display power and communication overheads that ultimately constrain system designers working within limited form-factor requirements.

To put this in context, the total required resolution under the 1D slice of ADF represented by our 2D plots is roughly 325 cycles (or 650 pixels) over 80$^\circ$. A brute force solution to this same 1D slice would require 30 cpd (60 ppd) resolution for the full 80$^\circ$, or 2400 cycles in any 2D slice. Thus this brute force solution obtains a 13.5\% RDF efficiency. This means that even if a uniform resolution 4k (8Mpx) display was used to implement the display, the total effective resolution could (theoretically) be driven through an interface that only supported 720p video bit rates.

\subsection{Stereoscopic and Multi-viewer Foveation}
While we have primarily discussed foveation in the context of a single viewer and even a single eye due to the head mounted display context, it is relatively straightforward to extend this classification to situations where more than one eye or viewer is viewing a single display. In cases where each eye is given an individualized view of the display content, the gaze direction can be thought of as a single, personalized fixation point. However, if a large display, such as a theater sized screen or shared public display \cite{dietz2019adaptive}, is intended to be foveated by multiple viewers simultaneously, it is beneficial to create one display foveal inset per viewer and steer them for each user to provide the optimal experience. Design considerations will require building in some practical limit to the number of users. As of this writing the most flexible approach would be to create a brute force light field display and foveate the content in software, though this is still to some extent limited by the required rendering throughput.

\subsection{Light Field and Variable Focus Displays}
In more complex displays (particularly those capable of displaying multi-focal content) the gaze direction may be 3 dimensional, as it includes the depth of gaze in addition to gaze direction  and is sometimes referred to as a \emph{fixation point}. In this case the display RDF needs to be extended to include an additional dimension characterizing the performance of the display across focal depth in addition to eccentricity.

Furthermore, each eye has its own unique gaze direction as defined previously, though typically the two eyes operate in concert to converge on particular objects. In these cases sensing the gaze direction for each eye independently and combining this information to gain more information about the user's fixation point within a given scene may present distinct advantages (namely depth information) over treating the two eyes as distinct, foveated entities.

For display systems providing focal cues (e.g. correct optical defocus blur), foveation can be applied to reduce the rendering complexity and/or to correct optical errors in the system. In both additive \cite{Lee2017Foveated} and multiplicative \cite{sun2017perceptually} near-eye light field displays, it has been reported that foveated rendering can decrease the number of rendered rays without compromising perceptual quality. Efficient foveated rendering is especially important in light field displays because of the large number of ``unused views'' or angular pixel components not seen by a viewer, for which rendering computations can be avoided. 

\subsection{Aliasing and Temporal Effects}
Typically aliasing in computer graphics is thought of as an image space artifact where jaggies show up spatially. 
Aliasing happens most often when the number of visibility or shading samples within a given pixel is small. 
To mitigate these artifacts, a variety of algorithms exist that reduce aliasing by increasing the number of samples used to reconstruct the final color within each pixel.
Aliasing can also happen in time, thus modern antialiasing algorithms \cite{karis2014high} often include some form of temporal reprojection and reconstruction.
Furthermore, while effects like crowding are somewhat understood, much of human peripheral vision, particularly the temporal effects on crowding in the periphery, are much more poorly understood. Common consensus is that the display will need to exceed at least 75-90hz, but it is not well understood what effect higher framerates and more continuous temporal display updates may have on the user.

\subsection{Contrast and Brightness}
It is well known that contrast of a displayed image is an essential factor in the human visual system's perception of resolution. Furthermore, without sufficient brightness, the human visual system switches from photopic to scotopic vision, further reducing resoltuion. We, mirroring the foundational work in human vision acuity, consider the ADF as defined relative to a reasonable contrast/brightness level for Snellen acuity measurement. Therefore the proposed taxonomy should only be applied for displays with sufficient brightness and contrast ratios. Future work may find ways to expand this taxonomy to cover low light and low contrast displays, though we believe the most valuable type of display to classify to assume normal light and contrast levels (photopic vision). 

\section{Conclusions}
This work is intended as an initial step towards a robust classification system for head-mounted (and otherwise) foveated displays. Though it is not without its limitations, we consider the combination of a gaze  and resolution contingent classification a promising direction for evaluation of future designs. Notably, our classification-based exploration of existing hardware designs reveals few Class 1, 2 or 3 designs, though recent research and product development suggests more will appear in the coming years. As more designs integrating eye tracking and active steering are introduced to the market, gaze-continence will likely become more dominant in the narrative of foveation. Until then, ADF and RDF comparison provides a useful basis for continued evaluation of foveal resolution designs.
We hope that this article will encourage the industry to come together to standardize terminology and classification as it relates to foveated displays. 

\acknowledgments{
This work was motivated by discussion among many colleagues including Alexander Majercik, Anjul Patney, Joohwan Kim, Mark Kilgard, Morgan McGuire, Peter Shirley, Turner Whitted, Ward Lopes, Kishore Rathinavel, Praneeth Chakravarthula, David Dunn, Henry Fuchs and Fu-Chung Huang.
}

\bibliographystyle{abbrv-doi}

\bibliography{related}

\end{document}